\documentclass{amsproc}

\newtheorem{theorem}{Theorem}[section]
\newtheorem{lemma}[theorem]{Lemma}

\theoremstyle{definition}

\newtheorem{corollary}[theorem]{Corollary}

\theoremstyle{remark}
\newtheorem{remark}[theorem]{Remark}

\numberwithin{equation}{section}

\begin{document}

\title{Necklaces with interacting beads: isoperimetric problems}

\author{Pavel Exner}
\address{Department of Theoretical Physics, Nuclear Physics
Institute, Academy of Sciences, 25068 \v{R}e\v{z} near Prague,
Czechia}
%\curraddr{}
\thanks{The research was supported by ASCR and its
Grant Agency within the projects IRP AV0Z10480505 and A100480501.}

%    General info
\date{June ??, 2005 and, in revised form, ????.}



\begin{abstract}
We discuss a pair of isoperimetric problems which at a glance seem
to be unrelated. The first one is classical: one places $N$
identical point charges at a closed curve $\Gamma$ at the same
arc-length distances and asks about the energy minimum, i.e. which
shape does the loop take if left by itself. The second problem
comes from quantum mechanics: we take a Schr\"odinger operator in
$L^2(\mathbb{R}^d),\; d=2,3,$ with $N$ identical point interaction
placed at a loop in the described way, and ask about the
configuration which \emph{maximizes} the ground state energy. We
reduce both of them to geometric inequalities which involve chords
of $\Gamma$; it will be shown that a sharp local extremum is in
both cases reached by $\Gamma$ in the form of a regular (planar)
polygon and that such a $\Gamma$ solves the two problems also
globally.
\end{abstract}

\maketitle

%%%%%%%%%%%%%%%%%%%%%%%%%%%%%%%%%%%%%%%%%%%%%%%%%%%%%%%%%%%%%%%%%

\section{Introduction}

Isoperimetric problems are certainly one of the topics which keeps
appearing in mathematical physics regularly. In the present paper
we are going to discuss two new examples of this kind. It may seem
that these problems differ mathematically and also have a
different physical background; we will show nevertheless that they
can be reduced to the same geometric question. A common feature is
that they both concern extremal properties of interaction between
$N\ge 2$ points placed at identical arc-length distances along a
closed curve $\Gamma$ of a fixed length $L>0$.

The first problem comes from classical electrostatics and concerns
\emph{charged necklaces}. Let $\Gamma:\:[0,L]\to\mathbb{R}^3$ be
such a loop and suppose that $N$ identical charges are placed at
the points $\Gamma(kL/N),\: k=0,1,\dots, N-1$. We ask about the
shape which this constrained family of point sources will take in
the absence of external forces, i.e. about \emph{minimum} of the
potential energy of the Coulombic repulsion.

The other problem comes from quantum mechanics. Having again a
loop $\Gamma$, now in two or three dimensions, we consider a class
of singular Schr\"odinger operators in $L^2(\mathbb{R}^d),\:
d=2,3,$ which are given formally by the expression
 % ------------- %
 \begin{equation} \label{formal}
 H_{\alpha,\Gamma}^N = -\Delta+\tilde\alpha \sum_{j=0}^{N-1}
 \delta\left(x-\Gamma\left( \frac{jL}{N}\right) \right)\,.
 \end{equation}
 % ------------- %
We will recall below how they can be defined properly, see
\cite{AGHH2}; following the terminology of this monograph we can
label the problem as a \emph{polymer loop}. This time we are
interested in the shape of $\Gamma$ which \emph{maximizes} the
ground state energy, of course, provided the discrete spectrum of
$H_{\alpha,\Gamma}^N$ is non-empty.

Our goal in this paper is to show that the two problems reduce
essentially to the same geometric question and that a sharp local
extremum is in both cases reached by the shape with a maximum
symmetry, in other words by a \emph{regular planar polygon} with
$N$ vertices denoted in the following as $\tilde{\mathcal{P}}_N$.
Furthermore, we will show that the regular polygon represents also
a \emph{global} solution to the problem by reducing the task to a
norm estimate of a particular operator on $\ell^2(\mathbb{Z})$.

The indicated quantum-mechanical isoperimetric problem was
formulated first in the paper \cite{Ex1}, to which we refer for a
deeper motivation, in the particular case when the curve $\Gamma$
was an equilateral polygon. The problem was restated there in
purely geometric terms and the existence of the maximizer was
proved locally. In \cite{Ex2} a similar local result was derived
for a continuous analogue of this problem in which the interaction
was supported by the entire curve $\Gamma$ and the maximizer is a
circle. The technical improvement in the last named paper was that
the geometric problem was viewed there from a general perspective
in terms of inequalities for chords of the loop, more
specifically, $\ell^p$ norms related to the functions
$\Gamma(\cdot+u)-\Gamma(\cdot)$. In the subsequent work \cite{EHL}
a simple Fourier analysis was used to show that in the
``continuous'' quantum-mechanical problem the circle is a
\emph{global} maximizer.

Our first aim here is to show first that the local proof of
\cite{Ex1} can be extended to the more general class of curves.
Then we will discuss how the argument of \cite{EHL} can be
modified to the present ``discrete'' situation. It appears that
the task is more involved than in the ``continuous'' case,
however, we will be able to reduce the question whether a regular
polygon $\tilde{\mathcal{P}}_N$ represents a global extremum in
both the isoperimetric problems described above to a well-defined
operator problem.

%%%%%%%%%%%%%%%%%%%%%%%%%%%%%%%%%%%%%%%%%%%%%%%%%%%%%%%%%%%%%%%%%

\section{Point interactions on a loop}

To begin with, let us make more precise our requirements on the
curve regularity. In what follows, we will suppose that for a
fixed dimension $d\ge 2$ \\ [.5em]
 % ------------- %
 \emph{\textbf{($\ell$)}} $\;\Gamma:\:[0,L]\to \mathbb{R}^d$ is a
 continuous, piecewise $C^1$ function such that $\Gamma(0)=\Gamma(L)$
 \phantom{\emph{\textbf{($\ell$)}} } and $|\dot \Gamma(s)|=1$ holds
 for any $s\in[0,L]$ for which $\dot \Gamma(s)$ exists. \\ [.5em]
 % ------------- %
The arc-length parametrization means in fact that we regard the
curve as a map $\mathbb{R}\to\mathbb{R}^d \;
\mathrm{(mod\,}L\mathrm{)}$. A shift in the argument is a trivial
reparametrization, so without loss of generality we may assume
that the point interactions are placed at
 % ------------- %
 \begin{equation} \label{sites}
 y_j:= \Gamma\left( \frac{jL}{N} \right)\,, \quad
 j=0,1,\dots,N-1\,;
 \end{equation}
 % ------------- %
the indices can be again regarded as integers, $y_j= y_{j
\mathrm{(mod\,}N\mathrm{)}}$. A distinguished element of the
described class is a \emph{regular polygon} for which the points
$y_j$ lie in a plane $\subset\mathbb{R}^d\:$ (this is trivial if
$d=2$) at a circle of radius $\frac{L}{N}\left( 2\sin
\frac{\pi}{N}\right)^{-1}$.

Let $Y_\Gamma:=\{y_j\,: j=0,\dots,N-1\}$ be the interaction
support. The object of our interest will the Hamiltonian
$-\Delta_{\alpha,Y_\Gamma}$ in $L^2(\mathbb{R}^d)$ with $N$ point
interactions, all of the same coupling constant
$\alpha\in\mathbb{R}$. It is defined conventionally through
boundary conditions which relate the generalized boundary values
at each site $y_k$, the coefficient at the singularity
(logarithmic for $d=2$, pole for $d=3$) and the next term in the
expansion -- see \cite{AGHH2} for a detailed discussion and recall
that the construction of a point-interaction Hamiltonian does not
work for $d\ge 4$. Recall also that $\alpha$ differs from the
formal coupling constant $\tilde\alpha$ in (\ref{formal}); it is
sufficient to realize that the absence of a point interaction
corresponds to $\alpha=\infty$.

It is obvious that that spectral properties of
$-\Delta_{\alpha,Y_\Gamma}$ and $-\Delta_{\alpha,Y_{\Gamma'}}$
corresponding to a pair of loops $\Gamma$ and $\Gamma'$ related
mutually by Euclidean transformations of $\mathbb{R}^d$ are the
same. This defines an equivalence relation on the set of all
loops; with a terminological abuse we will speak about curves
having in mind such equivalence classes. We will suppose that the
operator $-\Delta_{\alpha,Y_\Gamma}$ has a non-empty discrete
spectrum,
 % ------------- %
 \begin{equation} \label{nonempty}
 \epsilon_1 \equiv \epsilon_1(\alpha,Y_\Gamma):= \inf \sigma
 \left(-\Delta_{\alpha,Y_\Gamma}\right)<0\,,
 \end{equation}
 % ------------- %
which is true for any $\alpha\in\mathbb{R}$ if $d=2$, while in the
case $d=3$ it is a nontrivial assumption satisfied below a certain
critical value of $\alpha$ -- cf.~\cite[Sec.~II.1]{AGHH2}.

One of the main results of this paper referring to the second of
the problems mentioned in the introduction, the polymer loop, can
be formulated as follows:
 % ------------- %
 \begin{theorem} \label{loc_point}
 Assume \emph{\textbf{($\ell$)}} and (\ref{nonempty}); then
 $\epsilon_1(\alpha,Y_\Gamma)$ is for a fixed $\alpha$ and $L>0$
 locally sharply maximized by a regular polygon, $\Gamma=
 \tilde{\mathcal{P}}_N$.
 \end{theorem}
 % ------------- %

The proof will be done in several steps; in this section we will
demonstrated that the task can be reduced to a purely geometric
problem. With the usual notation, $k=i\kappa$ with $\kappa>0$, we
find the eigenvalues $-\kappa^2$ from the spectral condition,
 % ------------- %
 $$%\begin{equation} \label{spectral}
 \det Q_k =0 \qquad \mathrm{with} \qquad (Q_k)_{ij}:=
 (\alpha-\xi^k)\delta_{ij} - (1-\delta_{ij}) g^k_{ij}\,,
 $$%\end{equation}
 % ------------- %
where $g^k_{ij}:= G_k(y_i-y_j)$, or equivalently
 % ------------- %
 \begin{equation} \label{g}
 g^k_{ij} = \left\{ \begin{array}{ccc} \frac{1}{2\pi}
 K_0(\kappa|y_i-y_j|) &\quad\dots\quad& d=2 \\ [.3em]
 \frac{e^{-\kappa|y_i-y_j|}}{4\pi|y_i-y_j|}
 &\quad\dots\quad& d=3 \end{array} \right.
 \end{equation}
 % ------------- %
and the regularized Green's function at the interaction site is
 % ------------- %
 $$%\begin{equation} \label{xi}
 \xi^k = \left\{ \begin{array}{ccc} -\frac{1}{2\pi}
 \left(\ln\frac{\kappa}{2} +\gamma_\mathrm{E} \right)
 &\quad\dots\quad& d=2 \\ [.3em]
 -\frac{\kappa}{4\pi} &\quad\dots\quad& d=3 \end{array} \right.
 $$%\end{equation}
 % ------------- %
where $\gamma_\mathrm{E}$ is the Euler number. The matrix
$Q_{i\kappa}$ has $N$ eigenvalues counting multiplicity which are
decreasing in $(-\infty,0)$ as functions of $\kappa$ by \cite{KL,
AGHH2}. The spectral threshold $\epsilon_1(\alpha,\Gamma)$
corresponds to the point $\kappa$ where the lowest of the
indicated eigenvalues vanishes. Consequently, we have to check
that
 % ------------- %
 \begin{equation} \label{Gammaineq}
 \min \sigma(Q_{i\tilde\kappa_1}) < \min
 \sigma(\tilde{Q}_{i\tilde\kappa_1})
 \end{equation}
 % ------------- %
holds locally for $\Gamma\ne \tilde{\mathcal{P}}_N$. Here and in
the following the tilded quantities correspond always to regular
polygon, $\Gamma=\tilde{\mathcal{P}}_N$, in particular,
$-\tilde\kappa_1^2 = \epsilon_1(\alpha,\tilde{\mathcal{P}}_N)$.

Next we use the fact that the lowest eigenvalue of
$\tilde{Q}_{i\tilde\kappa_1}$ corresponds to the eigenvector
$\tilde\phi_1= N^{-1/2}(1,\dots,1)$, because by \cite{AGHH2} there
is a one-to-one correspondence between an eigenfunction $c=(c_1,
\dots,c_N)$ of $Q_{i\kappa}$ at the point, where the corresponding
eigenvalue vanishes, and the corresponding eigenfunction of
$-\Delta_{\alpha,Y_\Gamma}$ given by $c \leftrightarrow
\sum_{j=1}^N c_j G_{i\kappa}(\cdot-y_j)$, up to a normalization.
Again by \cite{AGHH2}, the principal eigenvalue of
$-\Delta_{\alpha,Y_\Gamma}$ is simple, so it has to be associated
with a one-dimensional representation of the corresponding
discrete symmetry group of $\tilde{\mathcal{P}}_N$; it follows
that the coefficients are the same, $c_1=\dots=c_N$. This yields
the expression
 % ------------- %
 \begin{equation} \label{minGammatilde}
 \min \sigma(\tilde{Q}_{i\tilde\kappa_1}) = (\tilde\phi_1,
 \tilde{Q}_{i\tilde\kappa_1} \tilde\phi_1) = \alpha -
 \xi^{i\tilde\kappa_1} - \frac{2}{N} \sum_{i<j}
 \tilde g_{ij}^{i\tilde\kappa_1}\,.
 \end{equation}
 % ------------- %
On the other hand, for the left-hand side of (\ref{Gammaineq}) we
have a variational estimate,
 % ------------- %
 $$%\begin{equation} \label{minGamma}
 \min \sigma(Q_{i\tilde\kappa_1}) \le (\tilde\phi_1,
 Q_{i\tilde\kappa_1} \tilde\phi_1) = \alpha -
 \xi^{i\tilde\kappa_1} - \frac{2}{N} \sum_{i<j}
 g_{ij}^{i\tilde\kappa_1}\,,
 $$%\end{equation}
 % ------------- %
which shows that it is sufficient to check validity of the
inequality
 % ------------- %
 \begin{equation} \label{Greenineq}
 \sum_{i<j} G_{i\kappa}(y_i-y_j) > \sum_{i<j}
 G_{i\kappa}(\tilde y_i-\tilde y_j)
 \end{equation}
 % ------------- %
\emph{for all} $\kappa>0$ and $\Gamma\ne \tilde{\mathcal{P}}_N$.
Let us introduce the symbol $\ell_{ij}$ for the chord length
$|y_i-y_j|$ and $\tilde\ell_{ij}:=|\tilde y_i- \tilde y_j|$, and
define the function $F:\: (\mathbb{R}_+)^{N(N-3)/2} \to\mathbb{R}$
by
 % ------------- %
 $$%\begin{equation} \label{F}
 F(\{\ell_{ij}\}): = \sum_{m=2}^{[N/2]}\: \sum_{|i-j|=m}
 \left[ G_{i\kappa}(\ell_{ij}) -G_{i\kappa}(\tilde\ell_{ij})
 \right]\,;
 $$%\end{equation}
 % ------------- %
we want to show that $F(\{\ell_{ij}\})>0$ except if
$\{\ell_{ij}\}= \{\tilde\ell_{ij}\}$. The function
$G_{i\kappa}(\cdot)$ is by (\ref{g}) \emph{convex} for a fixed
$\kappa>0$ and $d=2,3$, thus by Jensen's inequality we have
 % ------------- %
 $$%\begin{equation} \label{convF}
 F(\{\ell_{ij}\})\ge  \sum_{m=2}^{[N/2]} \nu_m
 \left[ G_{i\kappa}\left(\frac{1}{\nu_m} \sum_{|i-j|=m}
 \ell_{ij} \right) -G_{i\kappa}(\tilde\ell_{1,1+m}) \right]\,,
 $$%\end{equation}
 % ------------- %
where $\nu_n$ is the number of the appropriate chords,
 % ------------- %
 $$%\begin{equation} \label{ndiag}
 \nu_m:= \left\{ \begin{array}{ccl} N &\quad\dots\quad&
 m=1,\dots, \left[\frac{1}{2}(N-1)\right] \\ [.3em]
 \frac{1}{2}N &\quad\dots\quad& m=\frac{1}{2}N
 \quad\; \mathrm{for}\; N \;\mathrm{even} \end{array} \right.
 $$%\end{equation}
 % ------------- %
At the same time, $G_{i\kappa}(\cdot)$ is monotonously decreasing
in $(0,\infty)$, so the sought claim would follow if we
demonstrate the inequality
 % ------------- %
 \begin{equation} \label{lineq}
 \tilde\ell_{1,m+1} \ge \frac{1}{\nu_n} \sum_{|i-j|=m}
 \ell_{ij}
 \end{equation}
 % ------------- %
and show that it is sharp for at least one value of $m$ if
$\Gamma\ne \tilde{\mathcal{P}}_N$; in this way we have
accomplished the goal to reformulate the problem in geometric
terms.

%%%%%%%%%%%%%%%%%%%%%%%%%%%%%%%%%%%%%%%%%%%%%%%%%%%%%%%%%%%%%%%%%

\section{Inequalities for chord sums}

It is useful to discuss the geometric problem without dimensional
restrictions, i.e. for any $d\ge 2$. Denoting again for a given
$\Gamma$ by $y_j\in\mathbb{R}^d$ the points on the loop defined by
(\ref{sites}), we will study the following family of inequalities
 % ------------- %
 \begin{eqnarray}
 D_{L,N}^p(m):&\; \sum_{n=1}^N |y_{n+m}-y_n|^p\,
 \le\, \frac{N^{1-p}L^p \sin^p \frac{\pi m}{N}}{\sin^p \frac{\pi}{N}} \,,&
 \; p>0\,, \label{D+p} \\
 D_{L,N}^{-p}(m):& \sum_{n=1}^N |y_{n+m}-y_n|^{-p}\,
 \ge\, \frac{N^{1+p} \sin^p \frac{\pi }{N}}{L^p \sin^p \frac{\pi m}{N}}\,,&
 \; p>0\,, \label{D-p}
 \end{eqnarray}
 % ------------- %
for any $m=1,\dots, [\frac12 N]$, where $[\cdot]$ denotes as usual
the entire part; they contain as a particular case the
inequalities for equilateral polygons discussed in \cite{Ex2}.

The inequality (\ref{lineq}) is nothing else than $D_{L,N}^1(m)$.
In the next section we are going to demonstrate that it holds
locally, i.e. in the vicinity of $\tilde{\mathcal{P}}_N$, proving
thus Theorem~\ref{loc_point}. Furthermore, in Section~\ref{s:
global} we will show that a stronger claim can be made, namely
that the inequalities (\ref{D+p}) and (\ref{D-p}) hold generally
as long as $p\le 2$. Without imposing a restriction on $N$ this
seems to be an optimal result, because the example of a rhomboid
shows that $D_{L,4}^p(2)$ cannot be valid for $p>2$.

Notice that to establish the above described property one needs in
fact to prove the inequality $D_{L,N}^2(m)$ only as the following
simple result shows.
 % ------------- %
\begin{lemma} \label{auxil}
$D_{L,N}^p(m)$ implies $D_{L,N}^{p'}(m)$ if $p>p'>0$. Similarly,
$D_{L,N}^p(m)$ implies $D_{L,N}^{-p}(m)$ for any $p>0$.
\end{lemma}
 % ------------- %
\begin{proof} The first claim follows from convexity of $x\mapsto
x^\alpha$ in $(0,\infty)$ for $\alpha>1$,
 % ------------- %
 $$
 \frac{N^{1-p}L^p \sin^p \frac{\pi m}{N}}{\sin^p \frac{\pi}{N}} \ge
 \sum_{n=1}^N \left(|y_{n+m} -y_n|^{p'} \right)^{p/p'} \ge
 N \left( \frac1N \sum_{n=1}^N |y_{n+m} -y_n|^{p'} \right)^{p/p'}\,;
 $$
 % ------------- %
it is then sufficient to take both sides to the power $p'/p$. On
the other hand,
 % ------------- %
 $$ %\begin{equation} \label{Cerny}
 \sum_{n=1}^N |y_{n+m} -y_n|^{-p} \ge
 \frac{N^2}{\sum_{n=1}^N |y_{n+m} -y_n|^p} \ge
 \frac{N^{1+p} \sin^p \frac{\pi }{N}}{L^p \sin^p \frac{\pi m}{N}}
 $$ %\end{equation}
 % ------------- %
holds by Schwarz inequality; it gives the second claim and
completes the proof. \end{proof}

%%%%%%%%%%%%%%%%%%%%%%%%%%%%%%%%%%%%%%%%%%%%%%%%%%%%%%%%%%%%%%%%%

\section{Local extrema}

First we are going to prove Theorem~\ref{loc_point}. As we have
said one has to establish that the inequality (\ref{lineq}), or
$D_{L,N}^1(m)$, holds locally. Since the argument represents an
extension of the proof of Theorem~4.1 in \cite{Ex1} to a more
general class of constraints, we will skip details in repeating
the latter. We are looking for a local maximum of the function of
$Nd$ variables,
 % ------------- %
 $$%\begin{equation} \label{fm}
 f_m:\: f_m(y_1,\dots,y_N) = \frac{1}{N}\sum_{i=1}^N
 |y_i-y_{i+m}|\,,
 $$%\end{equation}
 % ------------- %
under the inequality-type constraints $g_i(y_1,\dots,y_n)\ge 0$,
where
 % ------------- %
 $$%\begin{equation} \label{gm}
 g_i(y_1,\dots,y_n):= \frac LN- |y_i-y_{i+1}|\,, \quad i=1,\dots,N\,;
 $$%\end{equation}
 % ------------- %
the true number of independent variables is $(N-2)(d-1)-1$ because
$2d-1$ parameters are related to Euclidean transformations and can
be fixed.

Following the convention for inequality-type constraints we
introduce slack variables $z_r,\: r=1,\dots,N,$ and Lagrange
multipliers $\lambda_r,\: r=1,\dots,N,$ which determine
 % ------------- %
 \begin{equation} \label{Km}
 K_m(y_1,\dots,y_N,z_1,\dots,z_N) := f_m(y_1,\dots,y_N) + \sum_{r=1}^N
 \lambda_r \left(g_r(y_1,\dots,y_n) -z_r^2 \right)\,.
 \end{equation}
 % ------------- %
The first thing to compute are the derivatives $\partial_{y_j}
K_m$ which are equal to
 % ------------- %
 $$%\begin{equation} \label{nabla}
 \frac{1}{N} \left\{
 \frac{y_j-y_{j+m}}{|y_j-y_{j+m}|}
 + \frac{y_j-y_{j-m}}{|y_j-y_{j-m}|} \right\}
 - \lambda_j N \frac{y_j-y_{j+1}}{L}
 - \lambda_{j-1} N \frac{y_j-y_{j-1}}{L}\,.
 $$%\end{equation}
 % ------------- %
Without loss of generality we may consider a regular polygon in
the plane spanned by the first two coordinate axes and to
parametrize its vertices in the following way,
 % ------------- %
 $$%\begin{equation} \label{regpar}
 \tilde y_{\pm m}= \frac LN \left( \pm\sum_{n=0}^{m-1}
 \cos\frac{\pi}{N}(2n+1), \sum_{n=0}^{m-1}
 \sin\frac{\pi}{N}(2n+1) \right)\,,
 $$%\end{equation}
 % ------------- %
so that
 % ------------- %
 $$%\begin{equation} \label{regdist}
 |\tilde y_j-\tilde y_{j\pm m}|= \frac LN \left[\left( \sum_{n=0}^{m-1}
 \cos\frac{\pi}{N}(2n+1) \right)^2 + \left(\sum_{n=0}^{m-1}
 \sin\frac{\pi}{N}(2n+1) \right)^2 \right] =: \frac{L\Upsilon_m}{N}\,.
 $$%\end{equation}
 % ------------- %
Hence the gradient components $\partial_{y_j} K_m$ will vanish for
$j=1,\dots,N$ provided we choose all the Lagrange multipliers in
(\ref{Km}) equal to
 % ------------- %
 \begin{equation} \label{lagrange}
 \lambda = \frac{\sigma_m}{N\Upsilon_m} \qquad \mathrm{with}\qquad
 \sigma_m := \frac{\sum_{n=0}^{m-1} \sin\frac{\pi}{N}(2n+1)}
 {\sin\frac{\pi}{N}} = \frac{\sin^2\frac{\pi m}{N}}
 {\sin^2\frac{\pi}{N}}\,;
 \end{equation}
 % ------------- %
notice that this quantity is always nonzero. At the same time, one
has to require vanishing of the derivatives
 % ------------- %
 $$%\begin{equation} \label{regdist}
 \partial_{z_j} K_m = 2\lambda_j z_j\,, \quad j=1,\dots,N,
 $$%\end{equation}
 % ------------- %
which means that at the extremum all the slack variables vanish,
$z_j=0$. This is not surprising; one naturally expects critical
points of the function $f_m$ to be reached under given constraints
with the neighbor distances maximal, i.e. for a polygon.

From this point on the argument proceeds as in \cite{Ex1}.
Evaluating the Hessian at the stationary point, one can reduce the
question about its negative definiteness to verification of the
inequalities
 % ------------- %
 \begin{equation} \label{chebysh}
 \sin\frac{\pi m}{N} \sin\frac{\pi r}{N} > \left|
 \sin\frac{\pi}{N} \sin\frac{\pi mr}{N} \right|\,, \qquad
 2\le r<m \le \left[\frac 12 N \right]\,,
 \end{equation}
 % ------------- %
or equivalently, the inequalities
$U_{m-1}\left(\cos\frac{\pi}{N}\right)> \left|U_{m-1}
\left(\cos\frac{\pi r}{N}\right)\right|$ for Chebyshev polynomials
of the second kind, which can be done directly.

The obtained result provides also a local solution to our
electrostatic problem.
 % ------------- %
 \begin{theorem} \label{loc_necklace}
 Under the assumption \emph{\textbf{($\ell$)}} the Coulomb energy
 of a charged necklace is locally sharply minimized by a regular
 planar polygon, $\Gamma = \tilde{\mathcal{P}}_N$.
 \end{theorem}
 % ------------- %
\begin{proof} For a given nonzero charge $q$ the potential energy equals
 % ------------- %
 $$ %\begin{equation} \label{real}
q^2\sum_{j\ne k} |y_j-y_k|^{-1} = q^2\sum_{m=1}^{\left[ \frac12
N\right]} \frac{\nu_m}{N} \sum_{n=1}^N |y_{n+m}-y_n|^{-1},
 $$ %\end{equation}
 % ------------- %
and since by Lemma~\ref{auxil} the inequality $D_{L,N}^1(m)$
implies $D_{L,N}^{-1}(m)$, the sum of all repulsion-energy terms
is locally sharply minimized by $\tilde{\mathcal{P}}_N$.
\end{proof}

%%%%%%%%%%%%%%%%%%%%%%%%%%%%%%%%%%%%%%%%%%%%%%%%%%%%%%%%%%%%%%%%%

\section{Global validity of mean-chord inequalities} \label{s:
global}

Let us look now what is needed to prove global validity of the
inequalities (\ref{D+p}) and (\ref{D-p}), in particular, to see
whether a Fourier analysis in the spirit of \cite{EHL} could help;
in view of Lemma~\ref{auxil} it is enough to consider the case
$p=2$ only. With the natural scaling properties in mind we may
without loss of generality put $L=2\pi$ and to express the
function $\Gamma$ through its Fourier series,
 % ------------- %
 \begin{equation} \label{expans}
 \Gamma(s) = \sum_{0\ne n\in\mathbb{Z}} c_n\, \mathrm{e}^{ins}
 \end{equation}
 % ------------- %
with $c_n\in\mathbb{C}^d$; since $\Gamma(s)\in\mathbb{R}^d$ the
coefficients have to satisfy the condition
 % ------------- %
 $$ %\begin{equation} \label{real}
 c_{-n} = \bar{c}_n\,.
 $$ %\end{equation}
 % ------------- %
The absence of the coefficient $c_0$ means naturally no
restriction; it can be always achieved by a choice of the
coordinate system.

It is convenient to impose slightly stronger regularity
requirements on $\Gamma$ assuming that it is of the $C^2$ class;
recall that validity of $D_{L,N}^2(m)$ can be extended from such a
family of loops to those satisfying the hypothesis
\emph{\textbf{($\ell$)}} by means of the Weierstrass theorem and
continuity of the functionals involved. In such a case
\cite[Sec.~VIII.1.2]{KF} the derivative of $\Gamma$ is a sum of
the uniformly convergent Fourier series
 % ------------- %
 \begin{equation} \label{der-expans}
 \dot\Gamma(s) = i \sum_{0\ne n\in\mathbb{Z}} nc_n\, \mathrm{e}^{ins}\,.
 \end{equation}
 % ------------- %
The assumed arc-length parametrization means $|\dot\Gamma(s)|=1$
giving thus the relation
 % ------------- %
 $$ %\begin{equation} \label{real}
 2\pi = \int_0^{2\pi} |\dot\Gamma(s)|^2\, \mathrm{d}s = \int_0^{2\pi}
 \sum_{0\ne l\in\mathbb{Z}}\: \sum_{0\ne n\in\mathbb{Z}} nl\, c^*_l\cdot c_n\,
 \mathrm{e}^{i(n-l)s} \,\mathrm{d}s\,,
 $$ %\end{equation}
 % ------------- %
where $c^*_l=(\bar c_{l,1},\dots, \bar c_{l,d})$ and dot marks the
inner product in $\mathbb{C}^d$, or equivalently
 % ------------- %
 \begin{equation} \label{norm}
 \sum_{0\ne n\in\mathbb{Z}} n^2 |c_n|^2 = 1\,.
 \end{equation}
 % ------------- %
Furthermore, using (\ref{expans}) we can rewrite the left-hand
side of $D_{2\pi,N}^2(m)$ as
 % ------------- %
 $$ %\begin{equation} \label{real}
 \sum_{n=1}^N\: \sum_{0\ne j,k\in\mathbb{Z}}\, c^*_j\cdot c_k\,
 \left(\mathrm{e}^{-2\pi imj/N}-1\right)
 \left(\mathrm{e}^{2\pi imk/N}-1\right)
 \mathrm{e}^{2\pi in(k-j)/N}
 $$ %\end{equation}
 % ------------- %
Next we change the order of summation and observe that
$\sum_{n=1}^N \mathrm{e}^{2\pi in(k-j)/N} = N$ if $j=k\;
(\mathrm{mod}\,N)$ and zero otherwise; this allows us to write the
last expression as
 % ------------- %
 \begin{equation} \label{chordsum}
 4N \sum_{l\in\mathbb{Z}} \sum_{\scriptsize{\begin{array}{c}0
 \ne j,k\in\mathbb{Z} \\ j-k=lN \end{array}}}\, |j|c_j^* \cdot
 |k|c_k  \left|j^{-1} \sin \frac{\pi mj}{N}\right|\,
 \left|k^{-1}\sin \frac{\pi mk}{N}\right| \,.
 \end{equation}
 % ------------- %
If $D_{2\pi,N}^2(m)$ should be valid, this quantity must not
exceed the right-hand side of (\ref{D+p}) for $p=2$; hence the
sought inequality is equivalent to
 % ------------- %
 \begin{equation} \label{ineq}
 \left( d,( A^{(N,m)}\otimes I )d\right) \le
 \left(\frac{\pi \sin\frac{\pi m}{N}}
 {N \sin\frac{\pi}{N}}\right)^2,
 \end{equation}
 % ------------- %
where the  vector $d\in \ell^2(\mathbb{Z})\otimes \mathbb{C}^d$
has the components $d_j:= |j|c_j$ and $A^{(N,m)}$ is an operator
on $\ell^2(\mathbb{Z})$ determined by its matrix representation as
 % ------------- %
 \begin{equation} \label{op_A}
 A^{(N,m)}_{jk} := \left\{ \begin{array}{ll}
 \left|j^{-1} \sin \frac{\pi mj}{N}\right|\,
 \left|k^{-1}\sin \frac{\pi mk}{N}\right| \quad & \mathrm{if}
 \quad 0 \ne j,k\in\mathbb{Z},\;\, j=k \:(\mathrm{mod}\,N) \\ [.7em]
 0 \quad & \mathrm{otherwise}
 \end{array} \right.
 \end{equation}
 % ------------- %
It is obvious that $A^{(N,m)}$ is bounded because its
Hilbert-Schmidt norm is finite. Since $d$ is a unit vector in view
of (\ref{norm}) we arrive at the following conclusion: the
inequality $D_{L,N}^2(m)$ with fixed $N,m$ is valid provided the
norm of the operator $A^{(N,m)}$ does not exceed the quantity at
the right-hand side of (\ref{ineq}).

This sufficient condition allows us to prove our second main
result.
  % ------------- %
 \begin{theorem} \label{global}
 Within the class of curves specified by the assumption
 \emph{\textbf{($\ell$)}} the inequalities $D_{L,N}^2(m)$, and thus
 also (\ref{D+p}) and (\ref{D-p}) with $p\le 2$ for fixed values
 of $N=2,3,\dots$ and $m=1,\dots, \left[\frac12 N\right]$, are valid.
 \end{theorem}
 % ------------- %
 \begin{proof}
For a given $j\ne 0$ and $d\in \ell^2(\mathbb{Z})$ the relation
(\ref{op_A}) gives
 % ------------- %
 $$ %\begin{equation} \label{}
 \left(A^{(N,m)} d\right)_j = \left|j^{-1} \sin \frac{\pi
 mj}{N}\right|\, \sum_{\scriptsize{\begin{array}{c}0
 \ne k\in\mathbb{Z} \\ k=j (\mathrm{mod}\,N) \end{array}}}\,
 \left|k^{-1}\sin \frac{\pi mk}{N}\right|\, d_k\,.
 $$ %\end{equation}
 % ------------- %
The norm $\|A^{(N,m)} d \|$ is then easily estimated by means of
Schwarz inequality,
 % ------------- %
 \begin{eqnarray*} %\label{}
 \|A^{(N,m)} d \|^2 &\!=\!& \sum_{0 \ne j\in\mathbb{Z}} j^{-2}
 \sin^2 \frac{\pi mj}{N}\;\: \Bigg|\!\!\!\!\sum_{
 \scriptsize{\begin{array}{c}0 \ne k\in\mathbb{Z} \\
 k=j (\mathrm{mod}\,N) \end{array}}}\,
 \left|k^{-1}\sin \frac{\pi mk}{N}\right|\, d_k\,\Bigg|^2 \\
 &\!\le\!& \sum_{n=0}^{N-1}\; \sin^4 \frac{\pi mn}{N}\; S_n^2 \,
 \sum_{\scriptsize{\begin{array}{c} n+lN\ne 0 \\ l\in
 \mathbb{Z} \end{array}}}\, |d_{n+lN}|^2, \phantom{AAAAAAAA}
 \end{eqnarray*}
 % ------------- %
where we have introduced
 % ------------- %
 $$ %\begin{equation} \label{}
 S_n:= \!\!\sum_{\scriptsize{\begin{array}{c} n+lN\ne 0 \\ l\in
 \mathbb{Z} \end{array}}}\, \frac{1}{(n+lN)^2} \,=\, \sum_{l=1}^\infty
 \left\{\frac{1}{(lN-n)^2} + \frac{1}{(lN-N+n)^2} \right\}\,.
 $$ %\end{equation}
 % ------------- %
However, the above series is easily evaluated to be
 % ------------- %
 $$ %\begin{equation} \label{}
 S_n= \left(\frac{\pi} {N \sin\frac{\pi n}{N}}\right)^2,
 $$ %\end{equation}
 % ------------- %
and since $\|d\|^2 = \sum_{n=0}^{N-1} \sum_{l\in\mathbb{Z}}
|d_{n+lN}|^2$, the sought inequality follows from (\ref{chebysh}).
\end{proof}

This allows us to strengthen our claims concerning the two
original problems.
 % ------------- %
 \begin{corollary} \label{glob_point}
 Adopt the assumptions \emph{\textbf{($\ell$)}} and (\ref{nonempty});
 then $\epsilon_1(\alpha,Y_\Gamma)$ is for a fixed $\alpha$ and $L>0$
 globally maximized by a regular polygon, $\Gamma=
 \tilde{\mathcal{P}}_N$.
 \end{corollary}
 % ------------- %
 \begin{corollary} \label{glob_necklace}
 Under the assumption \emph{\textbf{($\ell$)}} the Coulomb energy
 of a charged necklace is globally minimized by $\Gamma = \tilde{\mathcal{P}}_N$.
 \end{corollary}
 % ------------- %
 \begin{remark} \label{sharp}
 Notice that the inequality (\ref{chebysh}) used in the proof is
 sharp. That means that, in distinction to the ``continuous''
 analogue of our problem, the extremum cannot be reached in the
 class of $C^2$ smooth functions in which we have performed the
 described Fourier analysis.
 \end{remark}
 % ------------- %

%%%%%%%%%%%%%%%%%%%%%%%%%%%%%%%%%%%%%%%%%%%%%%%%%%%%%%%%%%%%%%%%%

\section{Concluding remarks}

Let us first comment on relations to the ``continuous'' case
treated in \cite{EHL} where the global validity of the
inequalities analogous to (\ref{D+p}) and (\ref{D-p}) was proved.
Notice that formally that situation corresponds to $N=\infty$. The
counterpart of the operator $A^{(N,m)}$ is then a multiple of the
unit operator and it is only necessary to employ the inequality
$\left| \sin \frac{\pi mj}{N}\right| \le \left|j\,\sin \frac{\pi
 m}{N}\right|$, or slightly more generally
 % ------------- %
 \begin{equation} \label{sin-ineq}
 \left| \sin jx\right| \le j\,\sin x
 \end{equation}
 % ------------- %
for any $j\in\mathbb{N}$ and $x\in(0,\frac12\pi]$, which is
checked easily by induction. In the present case with a finite $N$
the operator has infinitely many side diagonals such a simple
estimate based on (\ref{sin-ineq}) is too rough, because it yields
an unbounded Toeplitz-type operator, and one has to do better
using the matrix-element decay in (\ref{op_A}). Fortunately it can
be done as the proof of Theorem~\ref{global} shows.

Another aspect of the relation between the two cases is that in
the continuous case the analogue of (\ref{chordsum}) follows from
Parseval relation and the quantity is naturally invariant with
respect to shifts in the arc-length parametrization. This is not
true here; recall that the shift $s \to s+s_0$ is equivalent to
the replacement of $c_j$ by $c_j \mathrm{e}^{is_0}$, which changes
in general (\ref{chordsum}) due to the presence of the
off-diagonal terms.

Let us also comment briefly on various extensions of the present
problem restricting ourselves to the ``discrete'' situation only.
A natural question concerns the existence and properties of the
extrema in situations when we have no built-in symmetry, either by
assuming a nonconstant sequence of coupling parameters
$\{\alpha_j\}$ and/or taking their sites $s_j$ at the loop with a
non-equidistant distribution. In both cases the task becomes more
difficult because we can no longer use the relation
(\ref{minGammatilde}) which lead us to the geometric reformulation
based on the inequality (\ref{Greenineq}), in particular, the
solutions for the charged necklace and polymer loops will be now
in general different. Another extension could concern point
interaction family in $\mathbb{R}^3$ placed on a closed surface.
It is again not straightforward, however, because in distinction
to a curve such a surface cannot be locally rectified and the
answer will depend at the choice of the source sites at the
surface.

%%%%%%%%%%%%%%%%%%%%%%%%%%%%%%%%%%%%%%%%%%%%%%%%%%%%%%%%%%%%%%%%%

\bibliographystyle{amsalpha}

\end{document}